\documentstyle[12pt,epsfig]{article}

\input epsf

\newcommand{\frat}[2]{\frac{\textstyle #1}{\textstyle #2}}

\newcommand{\vf}[1]{\mbox{\boldmath $#1$}}

\begin{document}
\begin{center}
{\Large 
\bf Instanton liquid at finite quark density}\\
\vspace{0.5cm}
S. V. Molodtsov, G. M. Zinovjev$^\dagger$ 
\\
\vspace{0.5cm}
 {\small\it State Research Center,
Institute of Theoretical and Experimental Physics,
117259, Moscow, RUSSIA}\\ 
$^\dagger$
{\small \it
Bogolyubov Institute for Theoretical Physics,\\
National Academy of Sciences of Ukraine, 
UA-03143, Kiev, UKRAINE}
\end{center}
\vspace{0.5cm}
\begin{abstract}
The interaction of light quarks and instanton liquid in the phase
of nonzero values of chiral condensate is studied at finite
quark/baryon density. Calculating the generating functional in the
tadpole approximation we investigate the behaviour of dynamical quark mass 
and chiral condensate together with instanton liquid density (which shows 
insignificant increase) as a function of quark/baryon chemical
potential. We argue on the noticeable increase of quark density
at the point of developing the nonzero magnitude of diquark condensate
due only to the quark interaction with the instanton liquid.
\end{abstract}
\vspace{0.5cm}

PACS: 11.15 Kc, 12.38-Aw
\\
\\

\vspace{0.5cm}
\noindent
Since the discovery of instanton solutions in the Yang-Mills theory the 
functional integral language has become very popular and the most operational
instrument of quantum chromodynamics (QCD) phenomenology.
This formulation tolerates the nonperturbative identification of the field
configurations dominating the action and manifests the fundamental physics
features of QCD distinguishing it from other complex systems
with many degrees of freedom. In an overwhelming number of these theories
the quanta exchanged between interacting fermions according the initial
Lagrangian lead to the formulation of the concept of interaction potential and 
the definition of its concrete form. The dominating role of topological 
excitations of the gluon field and quantum oscillations about them in the QCD 
physics allowed us to understand and successfully explain new physical effects 
related, first of all, to the process of chiral symmetry breaking. Moreover, 
it compels us to have some doubts on the potential concept (perturbative 
description) pertinent.

In the intervening years considerable theoretical efforts were devoted
to extend the analytical description beyond dilute instanton gas approach
\cite{cdg}. It was shown that an effective instanton-anti-instanton 
interaction could stabilize the medium of these pseudoparticles (PP) \cite{2}
at the level of pretty reasonable values of major parameters
(average sizes and distances between PPs) forming rather instanton 
liquid \cite{sh}. The approach developed has succeeded in making interesting
(even quantitative) predictions although the lattice QCD studies only managed 
to put the problem of investigating the instanton role on the serious 
theoretical ground \cite{neg}. While much has been learned about the QCD
phase structure at finite temperature, this structure at nonzero 
quark/baryon densities has been significantly less explored because of
very serious technical difficulties, mainly, with putting the fermions (and
respective chemical potential) on the lattice. Nowadays, the possibility of
direct experimental measurements of instanton-induced effects is even discussed
in view of the future program for deep-inelastic scattering (small size 
instantons) at HERA \cite{rs}) and for ultrarelativistic heavy ion collisions 
at RHIC and LHC \cite{col}.  

This paper is addressed to investigate the light quark interactions with the 
instanton liquid (IL) in the phase of nonzero chiral condensate values at 
finite quark/baryon densities and continue the analysis we began in our 
preceding paper \cite{we0} where a new method was developed to estimate the 
IL back-reaction on the quark presence. Before it was always considered to be 
negligible. It is natural to start with the generating functional of the IL 
theory having the factorized form \cite{2} 
$$
{\cal Z}~=~{\cal Z}_g~\cdot~{\cal Z}_\psi~,
$$
where the factor ${\cal Z}_g$ originates information on the gluon condensate
while the fermion component ${\cal Z}_\psi$ provides the opportunity to study
the quark behaviour in the instanton medium and, in particular, the chiral and
diquark condensates together with their excitations \cite{2}. The factor 
${\cal Z}_g$ is treated in quasi-classical approximation supposing the
superposition of pseudo-particle (PP) fields $A_{\bar I I}(x;\gamma)$ which
are the Euclidean solutions of the Yang-Mills equations called the 
(anti-)instantons (${\bar I I}$) to be dominant saturating configurations. 
Such a solution is characterized by the parameters $\gamma={\rho,z,U}$ where  
$\rho$ denotes the PP size centered at the coordinate $z$ with the colour
orientation defined by the matrix $U$ in the space of colour group $SU(N_c)$. 
Let us suppose, for clarity, $N$ pseudo-particles available in four-volume $V$
and a liquid is topologically neutral $N_{\bar I}=N_I=N/2$. Formulating the 
variational maximum principle the 
following practical estimate has been obtained
$${\cal Z}_g~\simeq~e^{-\langle S\rangle}~,$$ 
with the average IL action $\langle S\rangle$ defined by the additive 
functional
\begin{equation}
\label{s}
\langle S\rangle=\int d z \int d\rho~n(\rho)~s(\rho)~,
\end{equation}
and with averaging the action per one instanton
\begin{equation}
\label{si}
s(\rho)=\beta(\rho)+5 \ln(\Lambda\rho)-\ln \widetilde \beta^{2N_c}
+\beta \xi^2\rho^2\int d\rho_1~n(\rho_1)\rho_1^{2}~,
\end{equation}
weighted with instanton size distribution function
\begin{equation}
\label{nrho}
n(\rho)=C~e^{-s(\rho)}=C~\rho^{-5} \widetilde\beta^{2 N_c} 
e^{-\beta(\rho)-\nu \rho^2/\overline{\rho^2}}~,
\end{equation}
where $\overline{\rho^2}=\int d\rho~\rho^2~n(\rho)/n=
\left(\frat{\nu}{\beta \xi^2 n}\right)^{1/2},~n=\int d\rho~n(\rho)=
\frat{N}{V}$,
$\nu=\frat{b-4}{2}$, $b=\frat{11~N_c-2~N_f}{3}$ and 
$N_f$ is the number of flavours.
The constant $C$ is defined by the variational principle in the selfconsistent 
way and
$\beta(\rho)=\frat{8\pi^2}{g^2}=-\ln C_{N_c}-b \ln(\Lambda \rho)$
($\Lambda=\Lambda_{\overline{MS}}=0.92 \Lambda_{P.V.}$) 
with another constant $C_{N_c}$ depending on the renormalization scheme,
in particular, here
$C_{N_c}\approx\frat{4.66~\exp(-1.68 N_c)}
{\pi^2 (N_c-1)!(N_c-2)!}$. The parameters $\beta=\beta(\bar\rho)$ and 
$\widetilde \beta=\beta+\ln C_{N_c}$ are fixed at the characteristic scale of 
the average PP size $\bar\rho$ . The constant 
$\xi^2=\frat{27}{4}\frat{N_c}{N_c^{2}-1} \pi^2$ 
gives information on interaction in the stochastic ensemble of PPs. Then 
the IL parameters extracted, for example, average PP size $\bar\rho$ and the IL
density $n$ agree reasonably with their values obtained from the 
phenomenological studies of the QCD vacuum.

In the quark determinant ${\cal Z}_\psi$ the quark fields are considered being 
{\it influenced} by  stochastic PP ensemble but the quark  
{\it backreaction} upon the instanton medium is usually ignored, i.e.
$${\cal Z}_\psi~\simeq~\int D\psi^\dagger D\psi~\langle\langle~
e^{S(\psi,\psi^\dagger,A)}~\rangle\rangle_A~.$$
Small values of the packing fraction parameter  
($n\bar\rho^4$) allow us to neglect the correlations between PPs and one
usually deals with the limit $N_c\to\infty$. The dominant contribution
to the action of fermion fields mainly comes from the zero modes 
$\Phi(x-z;\mu)$ which are the solutions of the Dirac equation 
$$[i\hat D(A_{\bar I I})-i\mu \gamma_4]~\Phi_{\bar I I}=0~,$$ 
with the chemical potential $\mu$ in the (anti-)instanton field
$A_{\bar I I}$, and their Weyl components 
\begin{center}
\vspace{0.25cm}
\parbox[b]{3.in}{$
\Phi(x;\mu)=
 \left 
 |\begin{array}{r}
\Phi_{\bar I}(x;\mu)\\
\Phi_{I}(x;\mu)
\end{array} \right 
|~,$}
\vspace{0.25cm}
\end{center}

\noindent
look like \cite{abr}
$$
\left[\Phi_{R,L}(x)\right]^{\alpha i}=\frat{\rho}{2\sqrt{2}\pi}~
e^{\mu x_4}\sqrt{\Pi(x)}~\partial_\mu
\left (\frat{e^{-\mu x_4} F(x;\mu)}{\Pi(x)}\right)
(\sigma^\pm)^i_{j}~\varepsilon^{jk}~U^\alpha_{k}~,
$$
with
$$\Pi(x)=1+\frat{\rho^2}{x_{4}^2+r^2}~,~~F(x;\mu)=\frat{1}{x_{4}^2+r^2}
\left[ \cos(\mu r)+\frat{x_4}{r}\sin(\mu r)\right]~.
$$
Here $\Phi_{R,L}=P_\pm~\Phi,~P_\pm=\frat{1\pm\gamma_5}{2}$, $r=|{\vf x}|$,
$\sigma^{\pm}_\mu=(\pm i{\vf \sigma},1),$  ${\vf \sigma}$ is the vector of
the Pauli matrices and $\varepsilon$ is the antisymmetric tensor. In 
particular, the quark determinant at $N_f=1$ might be written as \cite{2}
\begin{equation}
\label{6}
{\cal Z}_\psi~\simeq~\int D\psi^\dagger D\psi
~\left(\frat{Y^+}{VR}\right)^{N_+}
\left(\frat{Y^-}{VR}\right)^{N_-} 
\exp\left\{\int d x~\psi^\dagger(x)~(i\hat\partial-i\mu\gamma_4)
\psi(x)\right\}~,
\end{equation}
where the factor $R$ provides the dimensionless result. The preexponential
factors are responsible just for the instanton-induced interaction of quarks 
\cite{tH}
$$Y^{\pm}=i\int dz~dU~d\rho~n(\rho)/n
\int dxdy~\psi_{L,R}^\dagger(x)~(i\partial-i\mu)^{\mp}
\Phi_{\bar I I}(x-z)~
\widetilde\Phi_{\bar I I}(y-z)~(i\partial-i\mu)^{\pm}~\psi_{L,R}(y)~, 
$$
and the colour orientation averaging is performed by integrating over $dU$.
The designation $\widetilde \Phi$ is introduced for the conjugated zero mode
$\widetilde \Phi_{I}(x;\mu)=\Phi_{I}^\dagger(x;-\mu)$.
$\mu^\pm$ and similar designations are used for the four-vectors spanned on
the matrices $\sigma^\pm_{\nu}$, i.e. 
$\mu^\pm=\mu^\nu\sigma^\pm_{\nu}$, $\mu_\nu=({\vf 0},\mu)$.

Integrating over the auxiliary parameter $\lambda$ brings the quark 
determinant $Z_\psi$ to the exponential form convenient for the saddle 
point calculation 
\begin{eqnarray}
\label{9}
&&{\cal Z}_\psi\simeq\int \frat{d\lambda}{2\pi}~
\exp\left\{N\ln\left(\frat{N}{i\lambda VR}\right)-N\right\}
\times\nonumber\\
&&\times
\int D\psi^\dagger D\psi~\exp
\left\{\int \frat{dk}{(2\pi)^4}
\psi^\dagger(k)\left(-\hat k-i\hat\mu+
\frat{i\lambda}{N_c} \gamma_0(k;\mu)\right)\psi(k)\right\}~.\nonumber
\end{eqnarray}
The complex function $\gamma_0(k;\mu)$ of this part of the generating
functional is defined by the Fourier components of the zero modes 
$$\gamma_0(k;\mu)=(k+i\mu)_\alpha(k+i\mu)_\alpha~ h_\beta(k;\mu)
h_\beta(k;\mu)~,~~~
\Phi(k;\mu)^{i\alpha}=h_\mu(k;\mu)
(\sigma_\mu^{\pm})^i_{j}~\varepsilon^{jk}~U^\alpha_{k}~,
$$
\begin{eqnarray}
h_4(k_4,k;\mu)&=&\frat{\pi\rho^3}{4 k}
\{(k-\mu-ik_4)[(2k_4+i\mu)f^{-}_{1}+i(k-\mu-ik_4)f^{-}_2]+\nonumber\\
&+&(k+\mu+ik_4)[(2k_4+i\mu)f^{+}_{1}-i(k+\mu+ik_4)f^{+}_2]\}~,\nonumber
\end{eqnarray}
\begin{eqnarray}
h_i(k_4,k;\mu)&=&\frat{\pi\rho^3k_i}{4 k^2}
\left\{(2k-\mu)(k-\mu-ik_4)f^{-}_{1}+(2k+\mu)(k+\mu+ik_4)f^{+}_1+
\right.\nonumber\\
&+&\left[2(k-\mu)(k-\mu-ik_4)-\frat{1}{k}(\mu+ik_4)[k_4^{2}+(p-\mu)^2]
\right]f^{-}_{2}+\nonumber\\
&+&\left.
\left[2(k+\mu)(k+\mu+ik_4)+\frat{1}{k}(\mu+ik_4)[k_4^{2}+(p+\mu)^2]
\right]f^{+}_{2}\right\}~,\nonumber
\end{eqnarray}
where $k=|{\vf k}|$ for the spatial components of four-vector,
$$f_1^{\pm}=
\frat{I_1(z^\pm)K_0(z^\pm)-I_0(z^\pm)K_1(z^\pm)}
{z^\pm}~,
$$
$$f_2^{\pm}=\frat{I_1(z^\pm)K_1(z^\pm)}{z^2_{\pm}}~,
~~z^\pm=\frat{\rho}{2}\sqrt{k_4^{2}+(k\pm\mu)^2}~,$$
and $I_i,~K_i~(i=0,1)$ are the modified Bessel functions.

In order not to overload the following formulae with the unnecessary 
coefficients we introduce the dimensionless variables
$$
\frat{k_4\bar\rho}{2}\to k_4~,~~
\frat{k\bar\rho}{2}\to k~,~~\frat{\lambda\bar\rho^3}{2N_c}\to \lambda,
~~\psi(k)\to \bar\rho^{5/2}~\psi(k)~.
$$
Then it is easy to get the saddle point equation taking the 
form 
\begin{equation}
\label{per}
\frat{n\bar\rho^4}{\lambda}-2N_c\int\frat{dk}{\pi^4}
\frat{[\lambda^2 \gamma_{0}^2(k;\mu)]^{'}_\lambda}
{(k+i\mu)^2+\lambda^2 \gamma_{0}^2(k;\mu)}=0~,
\end{equation}
here the prime is attributed to differentiating in $\lambda$.

It has been shown in Refs. \cite{we0}, \cite{we} that the backreaction of 
quarks upon IL may be reproduced perturbatively with the small variations of 
the IL parameters $\delta n$ and $\delta \rho$ around their equilibrium values 
$n$ and $\bar\rho$. These variations are actually accommodated by the approach 
if the deformable (crumpled) (anti-)instantons of size $\rho$ being the 
function of $x$ and $z$, i.e. $\rho\to\rho(x,z)$ are utilized as the saturating
configurations. If the wave length of excitations is much larger than the
characteristic (anti-)instanton size $\bar\rho$ (for example, for $\pi$-meson)
the mean action per one instanton gains the extra term (looking like 
kinetic energy) generated by the deformable (anti-)instantons  
\begin{equation}
\label{estim}
\langle S\rangle\simeq\int d z \int d\rho~n(\rho)
~\left\{~\frat{\kappa}{2}~\left(\frat{\partial \rho}{\partial z}\right)^2
+s(\rho)\right\}~,
\end{equation}
where $\kappa$ is the kinetic coefficient being derived within the 
quasi-classical approach. If we strive for holding the precision declared in this
paper the kinetic coefficient should be fixed on a characteristic scale, 
for example, as $\kappa\sim\kappa(\bar\rho)$. Our estimates give the 
value of a few single instanton actions $\kappa\sim c~\beta$ 
(with the factor $c\sim 1.5$ --- $6$ depending on the  ansatz for the 
saturating configurations). Collecting the terms of the second order in 
deviation from the point of action minimum only  
$\left.\frat{d s(\rho)}{d\rho}\right|_{\rho_c}=0$, 
and taking approximately
\begin{equation}
\label{dec}
s(\rho)\simeq s(\bar\rho)+\frat{s^{(2)}(\bar\rho)}{2}~\varphi^2,~
\end{equation} 
where $s^{(2)}(\bar\rho)\simeq\left.\frat{d^2 s(\rho)}{d\rho^2}
\right|_{\rho_c}=\frat{4\nu}{\overline{\rho^2}},$ and the scalar field
$\varphi=\delta\rho=\rho-\rho_c\simeq\rho-\bar\rho$ is the field of deviations
from the equilibrium value of 
$\rho_c=\bar\rho~\left(1-\frat{1}{2\nu}\right)^{1/2}\simeq\bar\rho$,
it becomes clear the deformation field is described by the following Lagrangian
density
{\footnote{)
It seems to us the physical meaning of the deformation field is analogous to
the phonons of solid state physics. It is why we called them phononlike  
excitations of IL \cite{we}.}}): 
$${\cal L}=\frat{n\kappa}{2}
\left\{~\left(\frat{\partial \varphi}{\partial z}\right)^2+
M^2\varphi^2\right\}~,$$
with the mass gap of phononlike excitations  
$M^2=\frat{s^{(2)}(\bar\rho)}{\kappa}=
\frat{4\nu}{\kappa \overline{\rho^2}}$.
$M\approx 1.21~\Lambda$ 
for IL in the "quenched" approximation with the parameters fixed as $N_c=3$,
$~c=4$, $~\bar\rho~\Lambda\approx 0.37,~\beta\approx 17.5,~n~\Lambda^{-4}
\approx 0.44$  \cite{we}. 
 
Including the variations of zero modes in the quark determinant results in
changing those for
$\Phi_{\bar I I}(x-z,\rho)\simeq\Phi_{\bar I I}(x-z,\rho_c)+
\Phi_{{\bar I I}}^{(1)}(x-z,\rho_c)\delta\rho(x,z)$,
where $\Phi^{(1)}_{\bar I I}(u,\rho_c)=\partial\Phi_{\bar I I}(u,\rho)/\partial
\rho|_{\rho=\rho_c}$ and because of the adiabaticity constraint it is valid 
$\delta\rho(x,z)\simeq\delta\rho(z,z)=\varphi(z)$. The extra contributions
of scalar field generate the corrections in the kernels of factors $Y^{\pm}$ 
which are treated in the linear approximation in $\varphi$ and it is 
supposed $\rho_c\simeq\bar\rho$ everywhere. As result we have the 
effective Lagrangian with the Yukawa interaction of quarks and colourless
scalar field 
\begin{eqnarray}
\label{10}
&&{\cal Z}\simeq\int d\lambda~\widetilde{\cal Z}_g\int D\psi^\dagger D\psi~ 
D\varphi~\exp\left\{
N~\left(\ln\frat{n\bar\rho^4}{\lambda}-1\right)-\int 
\frat{dk}{\pi^4}~\frat{1}{2}~\varphi(-k)~
4~[k^2+M^2]~\varphi(k)\right\}\times\nonumber\\
[-.2cm]
\\[-.25cm]
&&\times\exp\left\{\int \frat{dkdl}{\pi^8}~\psi^\dagger(k)
~2~\left[\pi^4\delta(k-l)(-\hat k-i\hat\mu+i\lambda \gamma_0(k;\mu))+
i\lambda
\gamma_1(k,l;\mu)~\varphi(k-l)\right]~\psi(l)\right\}~.\nonumber
\end{eqnarray}
Let us mention the dimensionless variables as the scalar field
$\varphi(k)\to (n\kappa)^{-1/2}\bar\rho^3~\varphi(k)$ and its mass 
$\frat{M\bar\rho}{2}\to M$ were introduced here. $\widetilde{\cal Z}_g = 
e^{\langle S(s(\bar\rho))\rangle}$ denotes the part of the gluon component 
of the generating
functional which survives expanding the action per one instanton in small
deviation from the instanton equilibrium size. The general form of the kernel
$\gamma_1(k,l;\mu)$ is not of crucial importance for us here. We may simplify
calculation essentially because of the presence of quark condensate and
collect the leading contributions to the functional Eq. (\ref{10}) which are 
just the tadpole graphs. In this particular case the following purturbative
scheme (in the deviation from the condensate magnitude) is applicable
\begin{equation}
\label{scheme}
\psi^\dagger\psi~\varphi=\langle\psi^\dagger\psi\rangle~\varphi+
\{\psi^\dagger\psi-\langle\psi^\dagger\psi\rangle\}~\varphi~,
\end{equation}
and for the vertex we then have   
$$\gamma_1(k;\mu)=\frat{2}{(n\bar\rho^4\kappa)^{1/2}}\{
[k h+(k_4+i\mu)h_4][k\delta h+(k_4+i\mu)\delta h_4]+ 
[(k_4+i\mu) h-kh_4][(k_4+i\mu)\delta h-k\delta h_4]\}.
$$ 
Here we introduced the component $h$ spanned on the unit vector $\frat{k_i}{k}$
$h_i=\frat{k_i}{k}~h$ and $\delta h$ and $\delta h_4$ are produced by  
$h$ and $h_4$ while the substitutions $f^\pm_{i}\to g^\pm_{i}$ where 
$$g_1^{\pm}=f_1^{\pm}+
2(I_0^{\pm}K_0^{\pm}-I_1^{\pm}K_1^{\pm})~,~~  
g_2^{\pm}=-f_1^{\pm}-f_2^{\pm}~,
$$
done.

In the present paper two corrections are taken into account. One of them is
related to the changes of dynamical quark mass and chiral condensate due to
the interaction with the scalar field but another one comes from changing
the IL density invoked by the shift of the PP equilibrium size. The former
is given by the following contribution
\begin{eqnarray}
\label{12}
&&2~(i\lambda)^2
\int \frat{dkdl~dk'dl'}{\pi^{16}}~\gamma_1(k,l;\mu)~\gamma_1(k',l';\mu)
~~\psi^\dagger(k)\psi(l)~~\psi^\dagger(k')\psi(l')~
~\langle\varphi(k-l)~\varphi(k'-l')\rangle\simeq\nonumber\\
[-.2cm]
\\[-.25cm]
&&\simeq 4~\lambda^2
\int \frat{dk}{\pi^4}~\gamma_1(k;\mu)~\psi^\dagger(k)\psi(k)
~\int \frat{dl}{\pi^4}~\gamma_1(l;\mu)~{\mbox{ Tr}}~S(l)~D(0)~,\nonumber 
\end{eqnarray}
where the conventional definitions are introduced for the convolution of scalar
field
$$\langle\varphi(k)~\varphi(l)\rangle=
\pi^4\delta(k+l)~D(k)~,~~D(k)=\frat{1}{4~(k^2+M^2)}~,$$
and for the quark Green function $S(k)$
$$\langle\psi^\dagger (k)\psi(l)\rangle=
-\pi^4 \delta(k-l){\mbox{ Tr}}~S(k)~.$$ 
All the factors surrounding $\psi^\dagger(k)\psi(k)$ in Eq. (\ref{12}) 
give the extra contribution to the quark dynamical mass 
$\lambda\gamma_0(k;\mu)\to \lambda\Gamma(k;\mu)=
\lambda\gamma_0(k;\mu)+\lambda\delta \gamma_0(k;\mu)$,
where
\begin{equation}
\label{kirmas}
\delta \gamma_0(k;\mu)=\gamma_1(k;\mu)~(-2i\lambda)
~\int \frat{dl}{\pi^4}~\gamma_1(l;\mu)~{\mbox{ Tr}}~S(l)~D(0)~.
\end{equation}  
Let us introduce now the auxiliary function $c(\lambda)$ in order to 
emphasize the obvious parameter dependence of $\delta\gamma$, i.e. 
$$\delta \gamma_0(k;\mu)=\frat{N_c}{(n\bar\rho^4\kappa)^{1/2}}
~\frat{\kappa}{\nu}~\lambda^2 c(\lambda)~\gamma_1(k;\mu)~.$$
If the guark Green function is known $c(\lambda)$ could be defined by the
complete integral equation
\begin{equation}
\label{cl}
c(\lambda)=( n\bar\rho^4\kappa)^{1/2}~\int \frat{dk}{\pi^4}
~\gamma_1(k;\mu)~\frat{\Gamma(k;\mu)}{(k+i\mu)^2+\lambda^2 \Gamma^2(k;\mu)}~,
\end{equation}
what helps also to calculate $S(k)$.

It was shown in Ref. \cite{we0} the variation of equilibrium PP size
generates the IL density variation and can be given by the tadpole 
contribution $\varphi(0)~\Delta$ with
$\Delta=\frat{4 N_c}{n\bar\rho^4}~\lambda^2 ~c(\lambda)$
which produces an extra term in the mean action per one instanton
$$
\langle S\rangle=
\int d z~n \left\{\langle s\rangle-
\langle \Delta~\frat{\rho-\rho_c}{\bar\rho}\rangle\right\}~.
$$
The maximum variational principle helps to find the PP average size 
$\bar\rho\Lambda=\exp\left\{-\frat{2N_c}{2\nu-1}\right\}$
and to obtain the following equation to calculate the IL density \cite{we0}
\begin{equation}
\label{a7}
(n\bar\rho^4)^2-\frat{\nu}{\beta\xi^2}~n\rho^4=
\frat{\Delta}{\beta \xi^2}~
\frat{\Gamma(\nu+1/2)}{2\sqrt{\nu}~\Gamma(\nu)}~.
\end{equation}
Eventually, we come to the equation for the saddle point of generating 
functional Eq. (\ref{10}) in the form
\begin{equation}
\label{perper}
\frat{n\bar\rho^4}{\lambda}-
2N_c\int\frat{dk}{\pi^4}~\frat{2\lambda \Gamma^2(k;\mu)+\lambda^2 
\Gamma(k;\mu) \Gamma'(k;\mu)}
{(k+i\mu)^2+\lambda^2 \Gamma^2(k;\mu)}-(n\bar\rho^4)'~
\ln\frat{n\bar\rho^4}{\lambda}=0~,
\end{equation}
where the IL density variations are incorporated. The prime here signifies the 
derivative in  $\lambda$. But we need to know another derivative $c'$ to have 
the complete equation. With the definition of $c(\lambda)$ as given by 
Eq. (\ref{cl}), we obtain
$$(1-\lambda^2 A(\lambda))~
c'(\lambda)=2\lambda~A(\lambda)~c(\lambda)+B(\lambda)~$$
where the functions $A(\lambda)$ and $B(\lambda)$ are the following integrals
$$A(\lambda)=\alpha(\lambda)~\frat{N_c\kappa}{\nu}\int \frat{dk}{\pi^4}
\frat{\gamma^2_{1}(k;\mu)~[(k+i\mu)^2-\lambda^2 \Gamma^2(k;\mu)]^2}
{[(k+i\mu)^2+\lambda^2 \Gamma^2(k;\mu)]^4}~,
$$
$$B(\lambda)=-2\lambda~(n\bar\rho^4\kappa)^{1/2}~\int \frat{dk}{\pi^4}
\frat{\gamma_1(k)~\Gamma^3(k)}
{[(k+i\mu)^2+\lambda^2 \Gamma^2(k;\mu)]^2}~,
$$
and the factor $\alpha(\lambda)$ reads
$$\alpha(\lambda)=1-\lambda^2 \frat{N_c}{\beta\xi^2}
\frat{\Gamma(\nu+1/2)}{\nu^{1/2}\Gamma(\nu)}~
\frat{c(\lambda)}{n\bar\rho^4\left(n\bar\rho^4-
\frat{\nu}{2\beta\xi^2}\right)}~.
$$

The corresponding results for multiflavour approach $N_f>1$ could be obtained 
with the simple substitutions \cite{we0},\cite{dp2}. We have to introduce the
factor $N_f$ into the extra contribution to the dynamical mass in 
Eq. (\ref{12})
and to the factor $\Delta$ in the tadpole contribution. Besides, the logarithm 
in Eq. (\ref{perper}) should be modified according to the prescription 
$\ln\frat{n\bar\rho^4}{\lambda}\to 
\ln\left(\left(\frat{n\bar\rho^4}{2}\right)^{1/N_f}\frat{1}{\lambda}\right)$.
For clarity we take the factor $R$ to be equal one (see, for example, 
Eq. (\ref{6})) everywhere.  However, this parameter could be treated as the 
free one. Then the dimensional parameters of this approach are
given by the corresponding powers of $\Lambda_{QCD}$, for instance, 
$\rho\Lambda$ and could be selfconsistently obtained if the saddle point was 
already known.

\begin{figure}[htb]
\centerline{\epsfig{file=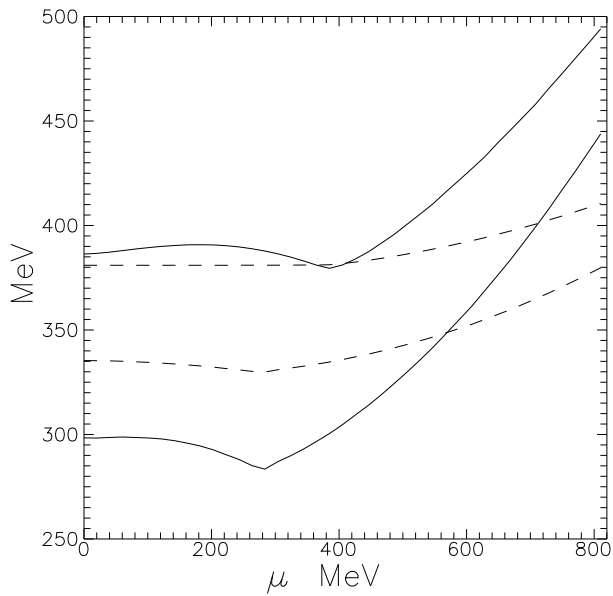,width=8cm}}
\vspace*{-.1 cm}
\caption{The dynamical quark masses (solid lines) and chiral condensates 
(dashed lines) as the functions of chemical potential $\mu$ at $\Lambda=280~MeV$
at  $N_c=3$, $N_f=2$. Both lower curves correspond to the calculation
with the modified generating functional.}
\end{figure}

In Fig. 1 we expose the calculation results with the positive root of Eq.
(\ref{a7})
$$n\bar\rho^4=\frat{\nu}{2\beta\xi^2}+
\left[\left(\frat{\nu}{2\beta\xi^2}\right)^2+
\frat{\Delta}{\beta \xi^2}~
\frat{\Gamma(\nu+1/2)}{2\sqrt{\nu}~\Gamma(\nu)}
\right]^{1/2}~,$$
for the dynamical quark mass $\lambda \Gamma(0,\mu)$ (solid lines) and for
the chiral condensate $-i\langle\psi^\dagger\psi\rangle$ (dashed lines) as 
the functions of chemical potential $\mu$ at $\Lambda=280~MeV$ and for $N_c=3$
, $N_f=2$. Both lower curves correspond to the calculation with modified 
generating functional when the tadpole contributions are taken into account. 
Apparently, the results obtained are well within the scope of QCD vacuum
phenomenology and, in principle, the perfect fit is achievable with slight
variation of $\Lambda_{QCD}$. 

Fig. 2 demonstrates rather insignificant change of the IL density when the 
quarks are in the phase where the chiral condensate develops non-zero values 
though the quark influence on IL is transformed into the small strengthening 
of gluon condensate manifesting itself in the supplementary attraction 
available in the system of quarks and gluons. These features fully agree with 
the corresponding results of Ref. \cite{rapp} where the IL behaviour was 
studied in the so-called cocktail model (with (anti-)in\-stan\-ton molecula 
admixed). In that model the IL density in the phase of broken chiral symmetry 
is almost constant up to the critical values of chemical potential.

\begin{figure}[htb]
\centerline{\epsfig{file=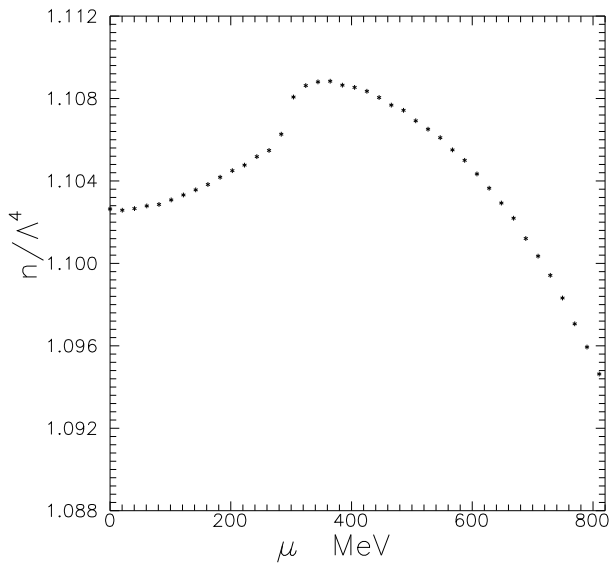,width=8cm}}
\vspace*{-.1 cm}
\caption{The IL density in the phase of nonzero chiral condensate values at
$N_c=3$, $N_f=2$ as function of chemical potential $\mu$.}
\end{figure}

The results of studying the saddle point parameter $\lambda$ (which is 
proportional to the free energy within the precision of one loop approximation)
at $N_c=3$, $N_f=2$ are shown in Fig. 3 where the upper curve was calculated  
without corrections coming from the quark feedback taken into account. In fact,
these results allow us to make one substantial qualitative prediction for 
the behaviour of critical chemical potential $\mu_c$ while transiting to the
colour superconductivity phase. Indeed, keeping in mind the results of Refs.
\cite{rapp}, \cite{diakcar} we should foresee the point $\mu_c$ shifting
perceptibly to the larger values because the parameter $\lambda$ for the
diquark phase at precritical values of $\mu$ is always disposed above than
for the corresponding curves of broken chiral symmetry phase. It means the  
crossing point with the lower curve is always situated further along the
$\mu$-axis than the crossing point with the upper curve. Such a situation
signals the density of quark matter in the critical interval could be 
comparable (or even larger) than normal nuclear density. In the meantime, we 
should remember that the IL model itself should obey some limitation for the 
chemical potential values. It is just the region where the transition to 
perturbative quark-gluon matter phase takes place. With quark matter density 
increasing the average interquark distance may be so small that the interquark 
gluon ('Coulomb') field strengths become comparable or even exceed the 
instanton ones. Then it is invalid to consider the (anti-)instanton ensemble 
as the saturating configuration and the IL approach is made irrelevant in this
region. 

\begin{figure}[htb]
\centerline{\epsfig{file=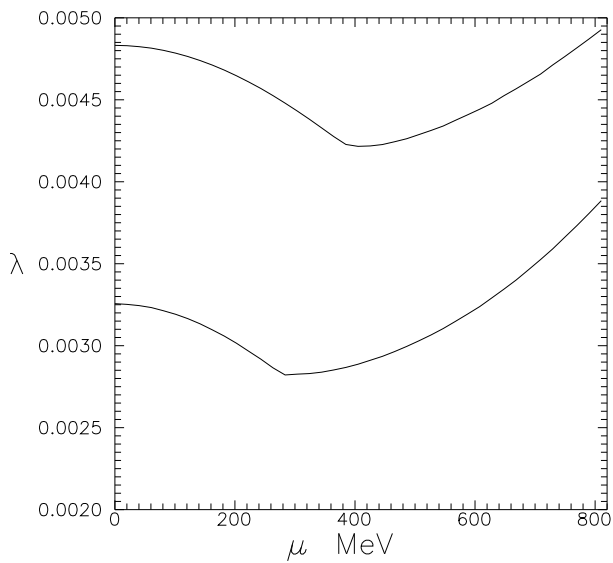,width=8cm}}
\vspace*{-.1 cm}
\caption{Saddle point $\lambda$  as function of $\mu$
in the phase of broken chiral symmetry 
at $N_c=3$, $N_f=2$. The upper curve 
corresponds to the solution with no perturbation of  IL.}
\end{figure}

\noindent
The authors are supported by STCU\#015c,
CERN-INTAS 2000-349, NATO~2000-PST.CLG 977482 grants. They are also 
indebted to the Faberge Fund for providing us with the excellent conditions 
for the completion of this paper.

\end{document}